\documentclass{aa}
\usepackage{graphicx}
\usepackage{txfonts}
\usepackage[colorlinks=true,linkcolor=blue,citecolor=blue]{hyperref}
\usepackage{natbib}
\bibpunct{(}{)}{;}{a}{}{,} % to follow the A&A style

\graphicspath{{./}{figures/}}
\graphicspath{{./}{figs/}}

\newcommand{\Ms}{$\textrm{M}_{\odot}$}

\newcommand{\kms}{$\textrm{km~s$^{-1}$}$}

\newcommand{\oi}{\,{\sc i}}
\newcommand{\ii}{\,{\sc ii}}
\newcommand{\iii}{\,{\sc iii}}
\newcommand{\Ha}{H$\alpha$}
\newcommand{\Hb}{H$\beta$} 
\newcommand{\Hg}{H$\gamma$}

\newcommand{\nb}{\textsc{NBursts}}

\begin{document}

   \title{Star formation in outer rings of S0 galaxies.}

   \subtitle{IV. NGC~254 -- a double-ringed S0 with gas counter-rotation}

   \author{Ivan Yu. Katkov
          \inst{1,2,3}
          \and
          Alexei Yu. Kniazev
          \inst{4,5,3}
          \and
          Olga K. Sil'chenko
          \inst{3}
          \and
          Damir Gasymov
          \inst{6,3}
          }

    \institute{New York University Abu Dhabi, PO Box 129188, Abu Dhabi, UAE
    \and
    Center for Astro, Particle, and Planetary Physics, NYU Abu Dhabi, PO Box 129188, Abu Dhabi, UAE
    \and
    Sternberg Astronomical Institute, M.V. Lomonosov Moscow State University, Universitetskiy pr., 13,  Moscow, 119992, Russia
    \and
    South African Astronomical Observatory, P.O. Box 9, Observatory, Cape Town, 7935, South Africa
    \and
    Southern African Large Telescope, P.O. Box 9, Observatory, Cape Town, 7935, South Africa
    \and
    Faculty of Physics, Moscow M.V. Lomonosov State University, 1 Leninskie Gory, Moscow 119991, Russia
    }

   \date{Received August 2, 2021; accepted December 6, 2021}

% \abstract{}{}{}{}{}
% 5 {} token are mandatory

  \abstract
  % context heading (optional)
  % {} leave it empty if necessary
   {}
  % aims heading (mandatory)
   {
   Although S0 galaxies are usually considered ``red and dead'', they often demonstrate star formation organized into ring structures. We aim to clarify the nature of this phenomenon and how it differs from star formation in spiral galaxies. 
   }
  % methods heading (mandatory)
   {
   We investigated the nearby, moderate-luminosity S0 galaxy NGC~254 using long-slit spectroscopy taken with the South African Large Telescope and publicly available imaging data.
    Applying a full spectral fitting, we analyzed gaseous and stellar kinematics as well as ionized gas excitation and metallicity and stellar population properties resolved by radius.
    An advanced approach of simultaneously fitting spectra and photometric data allowed us to quantify the fraction of hidden counter-rotating stars in this galaxy.}
  % results heading (mandatory)
  {We find that the ionized gas is counter-rotating with respect to the stars throughout NGC~254 disk, indicating an external origin of the gas.
  We argue the gas-rich galaxy merger from retrograde orbit as a main source of counter-rotating material.
   The star formation fed by this counter-rotating gas occurs within two rings: an outer ring at $R=55\arcsec - 70\arcsec$ and an inner ring at $R=18\arcsec$.
   The star formation rate is weak, 0.02 solar mass per year in total, and the gas metallicity is slightly subsolar.
   We estimated that the accretion of the gas occurred about 1~Gyr ago, and about 1\% of all stars have formed in situ from counter-rotating gas.

}
  % conclusions heading (optional), leave it empty if necessary
   {}


   \maketitle
%
%%%%%%%%%%%%%%%%% BODY OF PAPER %%%%%%%%%%%%%%%%%%

\section{Introduction}

Lenticular galaxies -- disk galaxies without spiral arms -- were introduced
by \citet{hubble} as a transition type between ellipticals and spirals. Their origin
is still vague; in particular, it is not clear why so many S0s lack noticeable star formation
while possessing a decent amount of cold gas \citep{pogge_esk93}. Also, unlike spirals,
lenticulars often demonstrate gas rotation decoupled from that of their stellar component
\citep{bertola92,kk_fisher96}, especially in a sparse environment \citep{atlas3d_10,isos0kin}; perhaps this phenomenon may cause star formation suppression \citep{s0_fp}.

A feature defining S0 as a morphological type is the presence of a large-scale stellar ring within
its disk structure \citep{devauc59}. In half of ringed S0s, the outer ring is bright in UV \citep{galex,kostuk15}, betraying recent star formation; in general, if a
gas-rich S0 proceeds current star formation, it is typically organized in ring-like structures \citep{salim12}.
We once noted that the star forming ring locations in S0s coincide with the radius where the
accreted gas lies in the main galactic plane \citep{s0_fp}. The net example of NGC~2551
proved that when the gas was confined to the stellar disk plane, it formed stars even though it was counter-rotating with respect to the stars.
In this paper, we present another S0 with star forming rings formed by counter-rotating gas -- NGC~254.

\begin{figure}
    \includegraphics[width=\columnwidth]{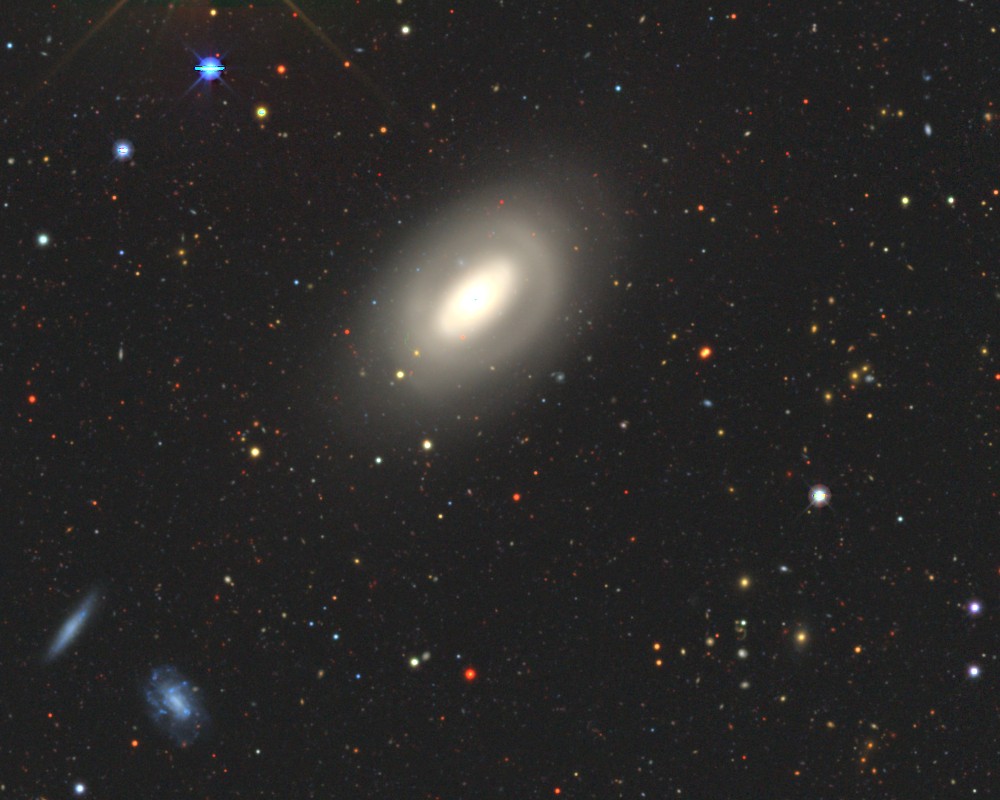}
    \caption{Galaxy NGC~254 and its environment, in composite colors, taken from the \href{https://www.legacysurvey.org/viewer?ra=11.8654&dec=-31.4226&layer=ls-dr9&zoom=13}{Legacy Survey image resource} \citep{Dey2019_legacysurveys}.}
    \label{fig:image}
\end{figure}

It is a nearby galaxy \citep[$D=17.1$~Mpc,][]{dist} classified as (R)SA(rl)$0^+$ by the NASA Extragalactic Database (NED)
and as S0/a with a bar and with a ring by Lyon-Meudon Extragalactic Database (HyperLEDA). It is a moderate-luminous galaxy ($M_B=-19.6$ based on the HyperLEDA data if corrected to the distance of 17.1~Mpc). As for the
environments, NGC~254 is formally attributed to the group of NGC~134 \citep{devauc75}, but there is a distance of 1.5~Mpc 
to this group center from NGC~254, and the only luminous galaxy in the vicinity of NGC~254 is the late-type NGC~289, 
in 400~kpc apart, beyond any gravitational tide. However, some small irregular satellites can be seen nearby 
(Fig.~\ref{fig:image}). NGC~254 is a rather gas-rich S0, with the HI mass of $3\times10^8$~\Ms\ \citep{hi}, or about 1--3 percent with respect to the total stellar mass. \citet{caldwell94} noted a dozen HII-regions in the inner ring of NGC~254, so it was a case of a star forming, gas-rich S0. 
\citet{buta_cat} included NGC~254 in his catalog of southern ringed galaxies, with the estimates of the outer stellar
ring major-axis radius of $62\arcsec$ and with the inner stellar ring major-axis radius of $25\arcsec$. In this
catalog, the galaxy was described as an unbarred one: ``a genuine and fine SA ringed galaxy'' \citep{buta_cat}. Despite the common view that rings in disk galaxies were phenomena related to bar resonances \citep{atha82}, in the catalog by \citet{buta_cat}, there were a lot of ringed S0 galaxies without obvious bars. In the recent statistics by \citet{diaz_knapen}, about a quarter of all ringed galaxies are classified as unbarred. A possible explanation is that the strong bars may be destroyed, leaving behind weak oval lenses in the centers of S0s. Ronald Buta has studied some of these galaxies in detail. As we demonstrate, below NGC~254 is morphologically similar to other unbarred double-ringed galaxies, such as NGC~7187 \citep{buta_n7187} and NGC~7702 \citep{buta_n7702}, which are characterized by a sharp blue inner ring and a diffuse reddish outer ring. 

We analyzed the photometric structure of of NGC~254 (Section~\ref{sec:structure}) and performed a spectral study to probe its kinematics, stellar populations properties, and ionized-gas localization and characteristics (Section~\ref{sec:spectral_study}). We discuss the star formation in NGC~254 and its ring nature in Section~\ref{sec:discussion} and make final conclusions in Section~\ref{sec:conclusions}.

\section{The structure of NGC~254}
\label{sec:structure}

Despite the visible regularity, the galaxy is classified very differently in numerous studies.
Among the large photometric surveys involving NGC~254, we could refer to the Carnegie-Irwin survey \citep{cair} and Spitzer Survey of Stellar Structure in Galaxies (S4G) \citep{S4G}.
The former covered the optical and near infrared (NIR) bands, from $B$ to $H$, and the latter was based on Spitzer space telescope data and explored the NIR bands of 3.6~$\mu$m and 4.5~$\mu$m. The morphological type of NGC~254 found in these surveys,
as well as listed in the extragalactic databases, disagrees radically: according to NED, Carnegie-Irwin, and \citet{buta_cat}, NGC~254 is
unbarred, (R)SA(rl)$0^+$, while HyperLEDA and S4G ascribe to it a large bar presence. \citet{buta_cat} found two stellar rings
in NGC~254, at $R=25\arcsec$ and $R=62\arcsec$. \citet{S4G_morph}, analysing the S4G data, found a bar with the semi-major axis of 27.3\arcsec\ and an inner lens with the semi-major axis of 30\arcsec. It is unclear which is correct.

\begin{figure}[hbt!]
\centering
    \includegraphics[width=\columnwidth]{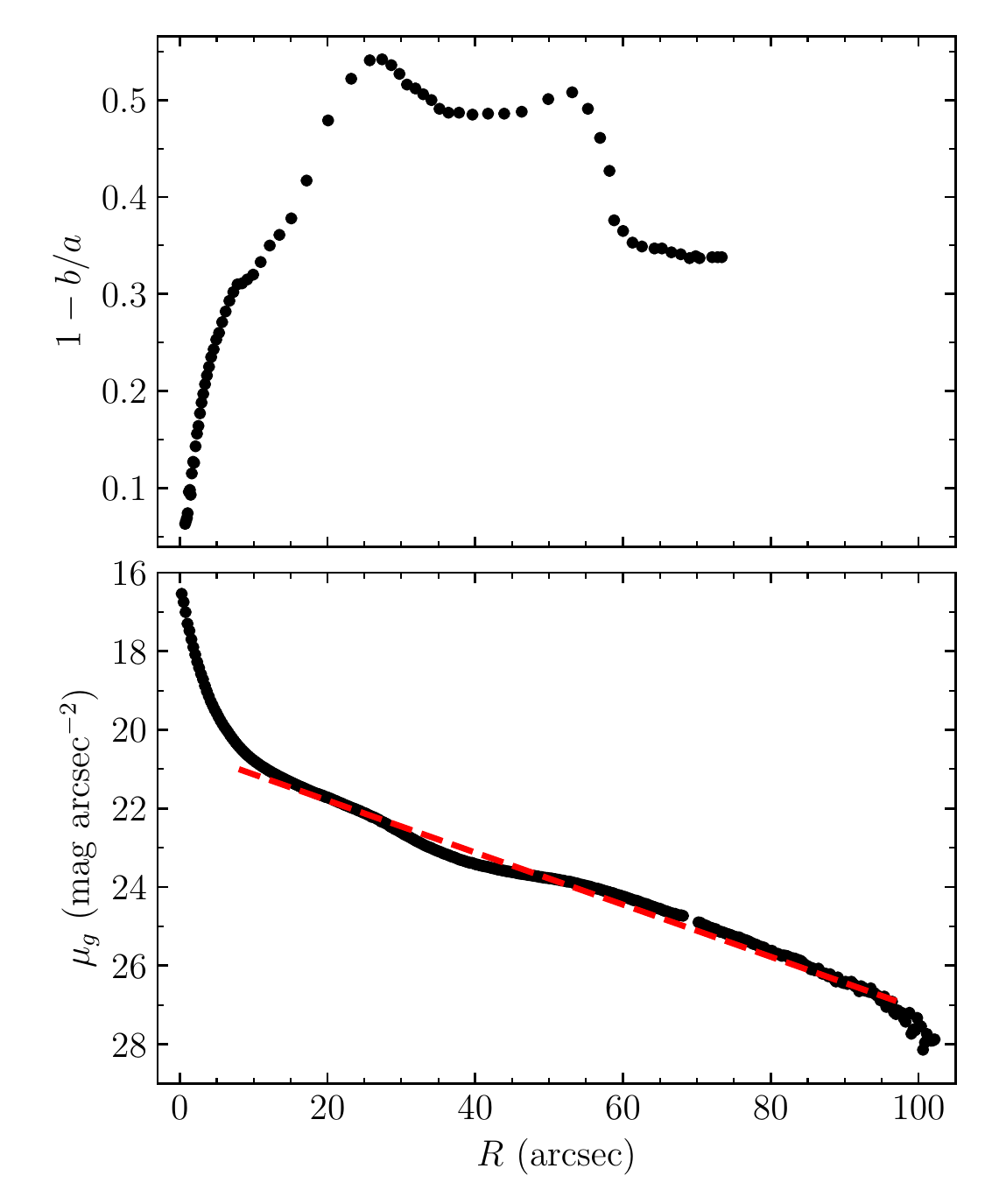}
    \caption{Isophote ellipticity radial profiles and the azimuthally averaged surface brightness profile in the $g$ band fit by an exponential law over the full disk-dominated area. Error bars are not shown because they are smaller than the size of the symbols.}
    \label{fig:profs}
\end{figure}

\begin{figure*}[hbt!]
\centering
    \includegraphics[width=\textwidth]{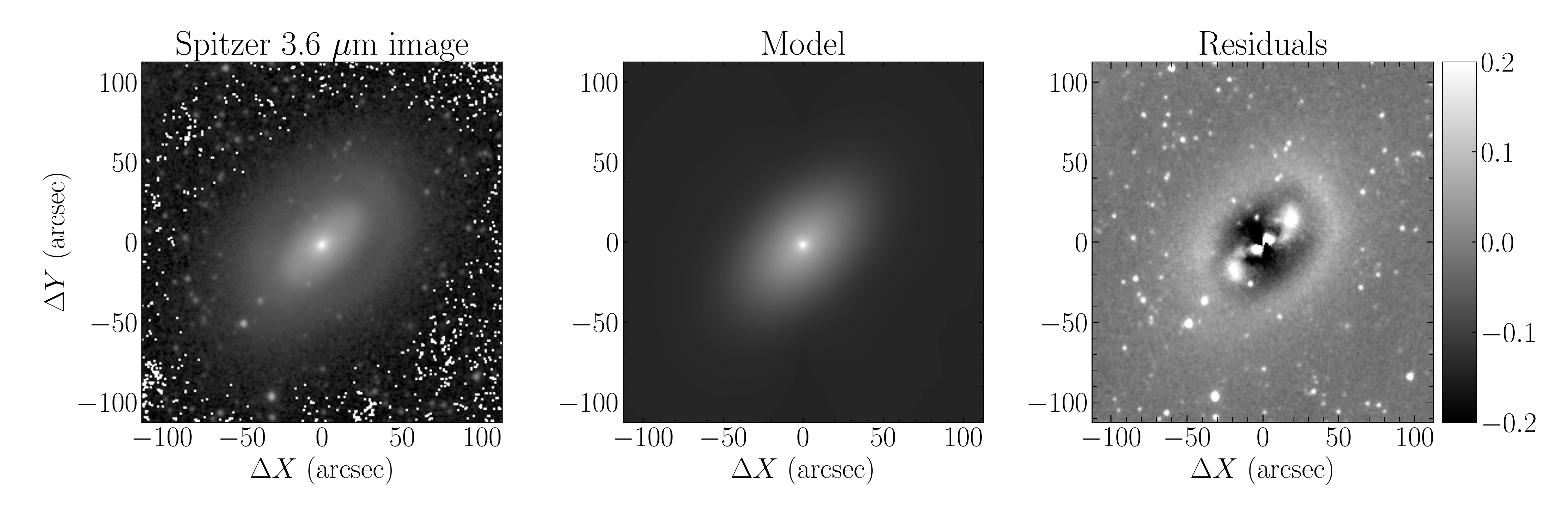}
    \caption{
    \textsc{galfit} modeling of 3.6~$\mu$m image of NGC~254. From left to right -- galaxy image, model, and residuals. We note that the galaxy image and model are shown in logarithmic scale, while residuals are given in the linear scale.}
    \label{fig:phot_galfit}
\end{figure*}

We inspected new photometric data available in The Dark Energy Camera Legacy Survey \citep[DECaLS,][]{Dey2019_legacysurveys}.
We retrieved the $g$-band image from the Legacy Survey website\footnote{ \href{https://www.legacysurvey.org/viewer?ra=11.8654&dec=-31.4226&layer=ls-dr9&zoom=13}{www.legacysurvey.org}} and constructed the azimuthally averaged surface brightness and isophote ellipticity radial profiles shown in Fig.~\ref{fig:profs}.
The major-axis position angle profile is not shown because it stays constant; the maximum deviation from the line of nodes is within three degrees and is observed just at the ring radii. Hence, the only possible morphological signature 
of a bar is the elevated isophote ellipticity between $R=15\arcsec$ and $R=60\arcsec$. However, two local maxima of the
isophote ellipticity correspond to both ring localizations -- at $R=25\arcsec$ and $R=55\arcsec$, and not to the bar end.
Since the stellar kinematics in the next section betrays the bar effect, as we show below, we would conclude that NGC~254 is an
unbarred galaxy despite the presence of two rings with a radius ratio between 2.2 and 2.6, which is what is predicted by the models of ring locations near the ultraharmonic $4 : 1$ and outer Lindblad resonances of a bar \citep{diaz_knapen}. Indeed, there is no observational difference between
ring radius ratio in barred and unbarred galaxies \citep{diaz_knapen}. Moreover, the disk profile classification
provided by \citet{S4G_disks}, Type II(R), seems doubtful because the $g$-band profile between $R=12\arcsec$ and $R=95\arcsec$ is divided into two segments, $12\arcsec -40\arcsec$ and $65\arcsec -95\arcsec$, which demonstrates similar exponential scale lengths and close central surface brightness in its inner and outer parts.
So, the whole disk can be described by a single exponential (radial scale $r_s=12.9$\arcsec), as is shown in Fig.~\ref{fig:profs}, with a very wide, diffuse outer stellar ring
visible between $R=54\arcsec$ and $R=70\arcsec$.

We also constructed a two-dimensional photometrical model by applying the \textsc{galfit} package \citep{Peng2002AJ....124..266P, Peng2010AJ....139.2097P} to a 3.6~$\mu$m image of NGC~254.
The model contains the S\'{e}rsic bulge, exponential disk, and sky background. 
The fitting procedure also uses point spread function (PSF), mask and uncertainty image provided in the S4G survey data.
Fig.~\ref{fig:phot_galfit} shows a 3.6~$\mu$m image, best-fit model, and residuals maps.
The radial scale of the modeled exponential disk $r_s=12.4\arcsec$ agrees very well with the scale determined from the radial brightness profile ($12.9\arcsec$).
The bulge component is very compact, the effective radius is $r_\mathrm{eff}=1.6\arcsec$, and it essentially corresponds to the core of the galaxy.
The residuals map clearly shows the outer ring at $R=55\arcsec$ and wavy ansae features similar to those in the galaxy NGC~7078 \citep[see Fig.~22 in][]{Buta2011arXiv1102.0550B}.

Bars are often associated with ansae, so 40\% of early-type barred galaxies also exhibit ansae features \citep{Martinez-Valpuesta2007AJ....134.1863M}.
Numerical simulations of barred galaxies show that the ansae features appear on the late phase of dynamical evolution of bars ($>5-7$~Gyr) \citep{Martinez-Valpuesta2006ApJ...637..214M, Long2014ApJ...783L..18L}.
Strong bars are not necessary for ansae forming, so we have an example of NGC~4151 \citep{Martinez-Valpuesta2007AJ....134.1863M} with ansae and an oval structure, which might be the product of the secular evolution of the bar \citep{Kormendy1979ApJ...227..714K}.
We also have an example of a lenticular galaxy, NGC~3998, where ansae features arise after bulge model subtraction from the image \citep{Laurikainen2009ApJ...692L..34L}. 
We tend to consider NGC~254 as a non-barred galaxy; however, it is clear that there is a non-axisymmetric central component, which may be an oval structure.
An important aspect is that this structure may play an important role in ring formation via resonances, as well as for the material transport to the galaxy center, which necessary for observed stellar population rejuvenation in the center (see panel \textit{e1} Fig.~\ref{fig:spresult}) and nuclear activity (see Fig.~\ref{fig:bpt2}).

\section{Results of spectral study}
\label{sec:spectral_study}

\subsection{Observations}
\label{sec:observations}

The long-slit observations of NGC~254 were made with the Robert Stobie Spectrograph
\citep[RSS;][]{Burgh03,Kobul03} of the Southern African Large Telescope
\citep[SALT,][]{Buck06,Dono06}. The observations were made on 12 June 2018 and 12 August 2018 with
the exposures of $1200\times 2$ seconds on each date, so the total exposure time was 80~minutes. 
A 1.25\arcsec\ slit width was explored under rather poor seeing conditions of $\sim 3\arcsec$. 
The slit was aligned with the isophote major axis in positional angle $PA=133^{\circ}$.
The grating PG900 was used to cover the spectral range of 4190$-$7260 \AA\ with a final reciprocal dispersion of $\approx0.97$ \AA\ pixel$^{-1}$ and spectral resolution of 5~\AA\  (full width at half maximum; FWHM).
The RSS pixel scale is 0.1267\arcsec, and we applied a binning factor of four to give a final spatial sampling of 0.507\arcsec~pixel$^{-1}$.
The slit length was approximately 8\arcmin, so spectra from the slit edges were used to
subtract the sky background. The primary reduction of the spectra was done as described in \citet{Kn08}.
We also corrected science spectra for scattered light in the instrument using the same approach as presented in \citet{Katkov2019MNRAS.483.2413K}.
Reference arc spectra were used to trace the instrumental resolution as a function of wavelength, which is a necessary ingredient of the spectra modeling.

\subsection{Spectral fitting}
\label{sec:sp_fitting}

To determine stellar population properties and emission-line parameters, we applied the \nb\ full spectral fitting code \citep{nburst_a, nburst_b}.
We fit the observed spectra with a simple stellar population models (SSP) built with the evolutionary synthesis code \textsc{pegase-hr} \citep{LeBorgne+04} based on the empirical stellar library Elodie3.1 \citep{Prugniel07}.
Each model template is a high-resolution spectrum interpolated from the model grid for a given pair of stellar age and metallicity, broadened with a Gaussian line-of-sight velocity distribution of stars (LOSVD).
To reduce the difference between the observed and template spectra, the model includes a multiplicative polynomial continuum on the order of 19.
Also, the model contains a set of strong emission lines (\Hg, \Hb, [O\iii], [O\oi], [N\ii], \Ha, [S\ii]) possessing the same Gaussian kinematics, which is, however, decoupled from the stellar kinematics.
Line fluxes are determined as a linear problem solving in every step within a nonlinear minimization loop providing a consistent solution.
The emission-line templates as well as stellar templates are built using the determined wavelength-dependent instrumental resolution.
This approach of simultaneous fitting the stellar continuum and emission lines with independent kinematics is similar to \textsc{gandalf} \citep{Sarzi2017ascl.soft08012S} or recent versions of the \textsc{ppxf} code \citep{Cappellari2017MNRAS.466..798C}.

\begin{figure*}
\centering
    \includegraphics[height=0.77\textheight]{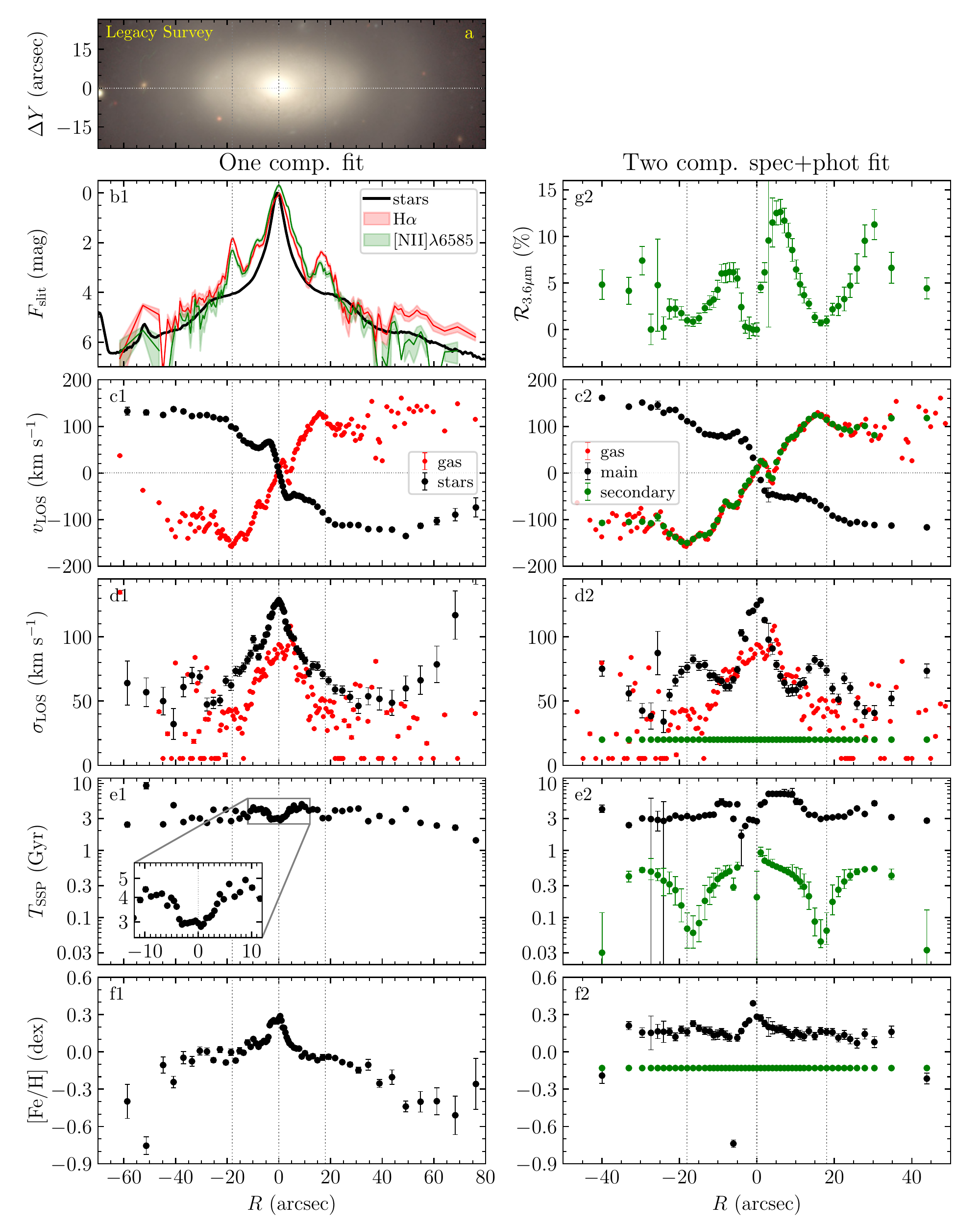}
    \caption{Results of spectrum fitting.
    The first column shows the result of a one-component fitting of the optical spectrum only (Section~\ref{sec:sp_fitting}), while the second column deals with a two-component spectro-photometrical fitting (Section~\ref{sec:spec_phot_fitting}).
    \textit{a)} Reference composite image manually built using $g$, $r$, and $z$-band data from DECaLS survey to avoid a saturated galaxy center.
    \textit{b1)} Stellar continuum level at 5500$\pm$20\AA\ in the rest frame along the slit (black line) and \Ha\ and [N\ii] fluxes (colored lines). For illustrative purposes, the stellar and emission-line profiles are converted to magnitudes assuming zero magnitude in the center.
    \textit{c1), c2)} Stellar and ionized gas line-of-sight velocity profiles.
    \textit{d1), d2)} Stellar and ionized gas line-of-sight velocity dispersion profiles.
    \textit{e1), e2)} SSP-equivalent age.
    \textit{f1), f2)} Stellar metallicity.
    \textit{g2)} Relative contribution of secondary component to 3.6~$\mu$m band.
    The black and green symbols in panels c2), d2), e2), and f2) show the parameters of the main and secondary stellar components in the spectro-photometric modeling.
    The vertical gray lines at $R=\pm18\arcsec$ correspond to the inner star forming ring.}
    \label{fig:spresult}
\end{figure*}

To obtain reliable radial profiles for stellar parameters, we binned a long-slit spectrum with a linearly increasing bin size, from 1 pixel in the galaxy center to 20 pixels (10\arcsec) at the outskirts, checking that the signal-to-noise ratio (S/N) in the stellar continuum at 4600$\pm$10\AA\ in every bin is be higher than 10.
For the emission-line profiles, we used a simple adaptive binning scheme that increases bin size to achieve the minimal required S/N=10 (total flux \Ha+[N\ii] after stellar continuum subtraction).
Fig.~\ref{fig:spresult} (panels in column 1) presents the resulting radial profiles.

The oldest galactic component is the bulge, with the mean luminosity-weighted age of more than 4~Gyr (panel \textit{e1} in Fig.~\ref{fig:spresult}).
At the same time, the stellar nucleus is younger, $T_\mathrm{SSP}\lesssim  3$~Gyr, and the outer disk at $R>40$\arcsec, including the outer ring, is also younger than 3~Gyr.
The stellar metallicity in the inner ring is close to the solar value, in the nucleus it is significantly higher (+0.3~dex), while the outer disk holds $\approx$3 times lower value (panel \textit{f1} in Fig.~\ref{fig:spresult}).

The extended distribution of the emission lines along the slit clearly reveals that the ionized gas counter-rotates relatively to the stars across the entire disk of the galaxy (panel \textit{c1} in Fig.~\ref{fig:spresult}).
The stellar velocity dispersion falls to 50~\kms, well below spectral resolution, corresponding to $V/\sigma \gtrsim 2$ around $R=20\arcsec-25\arcsec,$ which suggests a full prevalence of the dynamically cold disk at these radii.
In the $5\arcsec<|R|<15\arcsec$ range, the stellar velocity profile exhibits a flat section of smaller velocity amplitude.
This may be due to the dynamic effect of the bar-like oval structure aligned with the disk line of nodes \citep{miller_smith79}.
Another possibility is that we observe a small influence of the counter-rotating stellar component, which may also be the reason for the increase in velocity dispersion at $R>40\arcsec$.

\subsection{Looking for stellar counter-rotation}
\label{sec:loking_for_CR}

We tried to reveal signatures of counter-rotating stars by applying two different approaches.

\subsubsection{Non-parametrical LOSVD recovery}
\label{sec:nonparLOSVD}

\begin{figure}
\centering
    \includegraphics[width=\columnwidth]{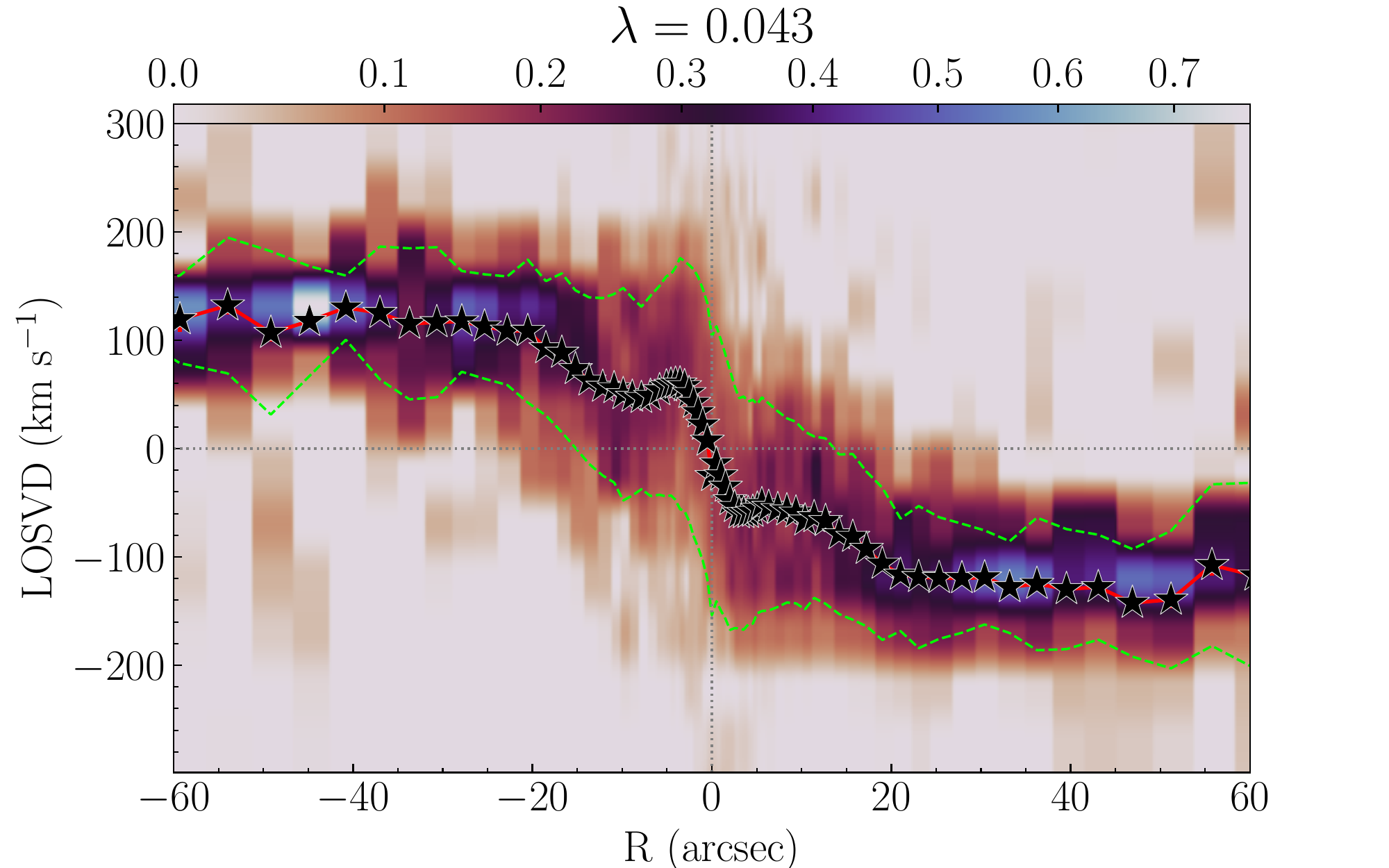}
    \caption{Non-parametrically recovered stellar LOSVD with regularization parameter $\lambda = 0.043$. Stellar velocity profile from an \nb\ analysis is shown with black stars as a reference. Dashed green lines correspond to $v_\mathrm{LOS} \pm \sigma_\mathrm{LOS}$ values.}
    \label{fig:stellar_losvd}
\end{figure}

Firstly, we recovered stellar LOSVD in a nonparametric manner using the technique developed and applied for revealing two-peak LOSVD structure in the counter-rotating galaxies IC~719 \citep{Katkov2013ApJ...769..105K}, NGC~448 \citep{Katkov2016MNRAS.461.2068K}, as well as for the recovery of LOSVD shape of NGC~7572 intending to distinguish a two-component (thin and thick) disk structure \citep{Kasparova2020MNRAS.493.5464K}.
In this approach, a galaxy stellar spectrum is considered as a result of convolution of a best-fit stellar spectrum found by \nb\ analysis with the non-parametrical stellar LOSVD.
The stellar LOSVD is determined by solving linear problem using regularization techniques (see details in \citet{Katkov2016MNRAS.461.2068K} and \citet{Kasparova2020MNRAS.493.5464K}).
Fig.~\ref{fig:stellar_losvd} shows the resulting reconstructed stellar LOSVD obtained using a first-order smoothing regularization, which means that we used the first order difference operator $D_1$ solving linear least-square problem minimizing $\chi^2 + \lambda \cdot D_1(\mathcal{L})$.

Despite numerous experiments with regularization parameters and a wavelength range, we found only weak and marginal signs of stellar counter-rotation, radically different from those observed in  stellar counter-rotating galaxies NGC~448, IC~719, and others.
There are some marginal spots at $R=-45\arcsec$, $v_\mathrm{LOS}\approx-100$~\kms and $R=-30\arcsec$, $v_\mathrm{LOS}\approx-70$~\kms.
In the flat region $5\arcsec<|R|<15\arcsec$, where parametric \nb\ fitting with a Gaussian LOSVD gives a smaller velocity amplitude, the nonparametric approach suggests an asymmetrical LOSVD shape.
Indeed, Fig.~\ref{fig:stellar_losvd} demonstrates an excess of positive velocities at $R>0\arcsec$ and an excess of negative velocities at $R<0\arcsec$, in the quadrants of the LOSVD panel where we would expect to see contribution from the counter-rotating stars.
However, LOSVD at $|R|<20\arcsec$ does not show a clear X-shaped structure, which is observed in galaxies hosting two stellar  counter-rotating disks (see Fig.~3 in \citet{Katkov2016MNRAS.461.2068K} and Fig.~3 in \citet{Katkov2013ApJ...769..105K}).
One possibility is that both disks have a high velocity dispersion, which washes out the zero-velocity dip.
In addition, due to the asymmetric drift effect, the high velocity dispersion disk rotates more slowly, making separation of a two-peaked LOSVD structure less prominent.
Another option is the bulge contribution at $|R|<20\arcsec$, but our two-dimensional \textsc{galfit} analysis suggests too compact a spherical component describing the galactic core to be noticeable at these radii.
Finally, this could be simply a bar/oval structure effect mentioned above.

We still have the question of why the large-scale counter-rotation structure (beyond $|R|=20\arcsec$) is not detected by the nonparametric LOSVD approach.
This could be the result of the actual absence of counter-rotating stars or the inapplicability of this approach to the case of NGC 254.
Due to signatures of ongoing star formation (see next section), we expect to see a young, stellar, counter-rotating component.
However, only one stellar template for both kinematics components is used in the explored technique.
In other words, our nonparametric technique is only capable of reconstructing a two-peak LOSVD structure if both components have comparable stellar population parameters and if both components contribute significantly to the integral light.
This may not be the case of NGC~254.

\subsubsection{Two-component spectro-photometrical fitting}
\label{sec:spec_phot_fitting}

The expected fraction of young stars may not contribute much to the optical spectrum, but it may become a main source of ultraviolet  (UV) emission.
Therefore, optical spectrum analysis combined with broad-band data including UV data can potentially help to reveal young counter-rotating stars on top of the older stellar population.
We used \textsc{NBursts+phot} mode of \nb\ in which optical spectra and broad-band spectral energy distributions (SED) are modeled simultaneously \citep{Chilingarian2012IAUS..284...26C, Grishin2021NatAs...5.1308G}.
For our spectral fitting, we used similar models to those described in Sect.~\ref{sec:sp_fitting}, while photometric models are computed with the \textsc{PEGASE.2}  code \citep{Fioc1997A&A...326..950F} using the low-resolution BaSeL synthetic stellar library \citep{Lejeune1997A&AS..125..229L}.
During minimization in the \textsc{NBursts+phot} mode, the total $\chi^2$ is equal to a sum of spectral and photometric contributions: $(1-\alpha)\chi^2_\mathrm{spec} + \alpha \chi^2_\mathrm{phot}$.
In our modeling, we used equal weights of the contributions $\alpha=0.5$.

\begin{figure*}
\centering
    \includegraphics[width=1\textwidth]{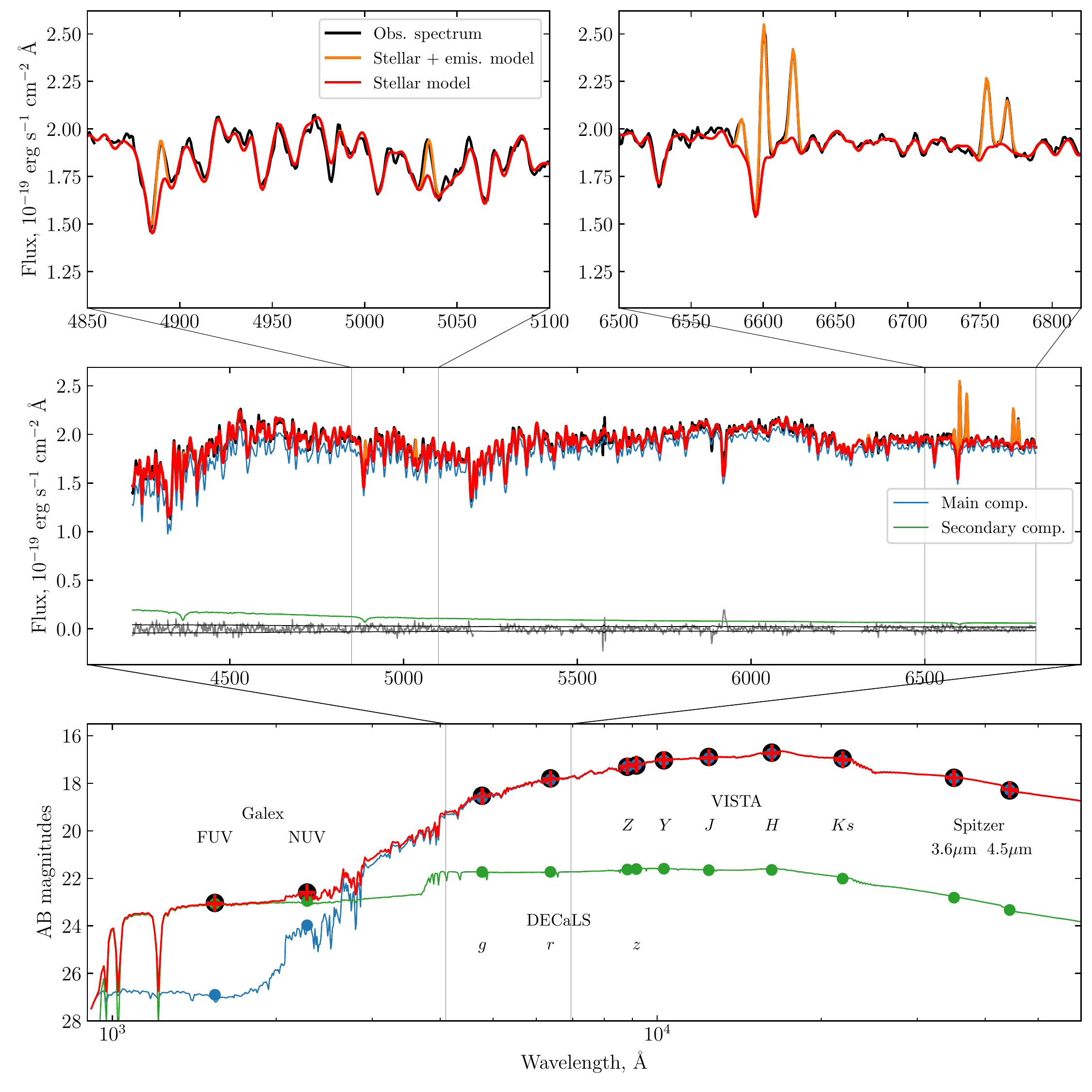}
    \caption{Result of two-component spectro-photometric modeling of the spatial bin at $R=+18\pm3\arcsec$. Middle and top panels show spectrum (black) and best-fit spectral model (red). Two stellar components of the model are shown in blue and green colors. Orange lines highlight the emission-line model. Residuals are presented in the mid panel in gray bordered by the error level (black lines). Bottom panel shows the SED part of the model. Observed magnitudes are shown in large black circles, and the best-fit SED model is shown with red crosses. Blue and green circles correspond to SED models of individual components whose low-resolution spectra are shown by colored lines.}
    \label{fig:spec_phot}
\end{figure*}

We compiled the SED dataset, which includes ultraviolet $FUV$, $NUV$ profiles from Galaxy Evolution Explorer (GALEX) taken from \citep{S4G_galex}, optical $grz$ data from The Dark Energy Camera Legacy Survey (DECaLS) survey, $ZYJHKs$ images from the VISTA Kilo-Degree Infrared Galaxy Survey (VIKING) retrieved through the ESO Archive Science Portal\footnote{\url{http://archive.eso.org/scienceportal/}} , and Spitzer 3.6~$\mu$m and 4.5~$\mu$m of the S4G Survey \citep{S4G} downloaded through the NASA/IPAC Infrared Science Archive\footnote{\url{https://irsa.ipac.caltech.edu/data/SPITZER/S4G/}}.
We estimated and subtracted sky background level by applying a $\sigma$-clipping technique\footnote{We used \textsc{sigma\_clipped\_stats()} function from the \textsc{astropy.stats} package: \url{https://docs.astropy.org/en/stable/stats/}.} for VISTA and Spitzer images, while DECaLS provides already sky-subtracted data.
Except for the GALEX data, where we used azimuthally averaged $FUV$, $NUV$ profiles from \citet{S4G_galex}, we extracted AB mag arcsec$^{-2}$ values along the long slit and then performed the same spatial binning for long-slit spectra and photometry profiles.

For SED analysis, it is important to take into account dust extinction. In \textsc{spec+phot} mode of \nb, $A_V$ is an external parameter. We applied $A_V=0.042$, determined based on a modeling of the full spectral energy distribution from ultraviolet to submillimeter in the frame of DustPedia photometry and the \textsc{cigale} code \citep{Bianchi2018A&A...620A.112B,Nersesian2019A&A...624A..80N}.
The galactic extinction $A_{V,\mathrm{gal}}=0.06$, drawn from the NED, is also taken into account.

\begin{figure}[hbt!]
\centering
    \includegraphics[height=0.76\textheight]{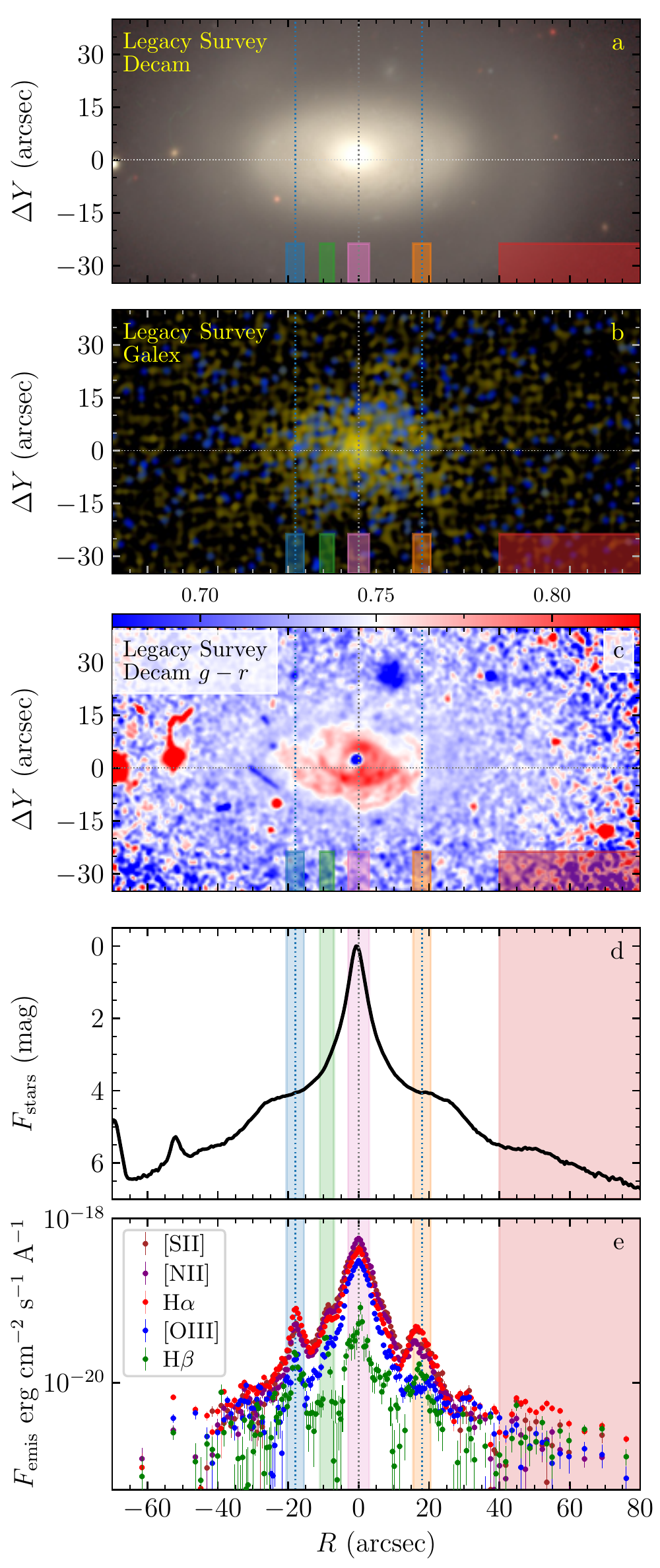}
    \caption{Comparison of imaging and spectral data.
    \textit{a)} Composite image built using $grz$ DECaLS data.
    \textit{b)} GALEX color image retrieved using the Legacy Survey \href{https://www.legacysurvey.org/viewer/cutout.jpg?ra=11.8649&dec=-31.4222&layer=galex&pixscale=0.50}{cutout service}.
    \textit{c)} $g-r$ color map constructed based on the DECaLS data.
    \textit{d)} Stellar continuum level at 5500$\pm$20\AA\ in the rest frame along the slit. The zero point is chosen to have zero magnitude at the center of the galaxy.
    \textit{e)} Emission line fluxes recovered from the long-slit spectra.
    Color stripes are shown for reference and comparison with other figures and represent emission-line peaks at the inner star forming ring $R=\pm18\arcsec$, external diffuse emission region at $R=45\arcsec-75\arcsec$, and the galaxy nuclear and emission enhancement at $R=-10\arcsec$.}
    \label{fig:spresult2}
\end{figure}

The secondary component has poorly constrained parameters due to its small contribution to the optical spectrum.
In addition, the spectrum of the young stellar population contains fewer absorption features, which further impairs parameter recovery.
It is difficult to determine the properties of the stellar population and especially kinematics of the secondary component just by SED fitting.
However, the relative weight can be well limited because the SED cannot be described by a single SSP component.
To stabilize the solution, we assumed that the secondary component inherits the kinematics of the ionized gas; so, during the fit both emission lines and secondary stellar component have identical velocities.
We also fixed the velocity dispersion at $\sigma_{\star,\mathrm{2comp}}=20$~\kms\ and stellar metallicity at -0.13~dex for the secondary component.
The motivation for the latter limitation is the measured metallicity of the ionized gas in the star forming inner ring equal to this value (see Section~\ref{sec:ionized_gas}).
All parameters of the main component, age, and weight of the secondary component are free fitting parameters.

To estimate uncertainties of the derived parameters, we used Monte Carlo modeling.
We generated 10,000 synthetic spectra for each spatial bin by adding to the best-fit model of the spectrum and SED a random noise in accordance with the error level in the original data.
Then, by analyzing these mock spectra in the same manner as the original data, we determined output parameters whose distributions allowed us to estimate statistical uncertainties of the model parameters.
The resulting parameter profiles are shown in the second column of Fig.~\ref{fig:spresult}.
Fig.~\ref{fig:spec_phot} illustrates spectro-photometrical analysis for one spatial bin at $R=+18\pm3\arcsec$.

Firstly, the analysis suggests a very young age of the secondary component at $|R|\approx18\arcsec$ (panel \textit{e2} in Fig.~\ref{fig:spresult}).
In Section~\ref{sec:ionized_gas}, we show that in this region we observe an enhancement of emission lines caused by the current star formation.
The blue ring at the same radius observed in the GALEX UV images (panel \textit{b} in Fig.~\ref{fig:spresult2}) suggests that we are dealing with a star-forming \textit{\emph{ring}}, despite the use of long-slit spectroscopy, which cannot reveal the two-dimensional nature of this structure.
Secondly, we note that the secondary component contributes $\approx1\%$ to the 3.6$\mu$m luminosity, that is stellar density, at $|R|\approx18\arcsec$ (panel \textit{g2} in Fig.~\ref{fig:spresult}).
Closer and farther than $|R|\approx18\arcsec$, the relative stellar weight of the secondary component $\mathcal{R}_\mathrm{3.6\mu m}$ increased to $10-15\%$.
However, its age also changed significantly from 50~Myr to 500~Myr (panel \textit{e2} in Fig.~\ref{fig:spresult}).
We interpret this effect to mean that at $|R|\approx18\arcsec$ we definitely detect a very \textit{\emph{young counter-rotating}} stellar component, as it is currently forming from gas that is counter-rotating to the main stellar disk.
Beyond this radius, young stars disappear, and we begin to detect an older population, $\approx500$~Myr old, whose contribution to the stellar density is more prominent.
We
cannot definitely conclude from our spectro-photometric analysis whether this part of the stellar population is counter-rotating or not.
We made various fits allowing the secondary component to have an individual velocity unrelated to the ionized gas, but experiments show a very high variation in the output velocity from bin to bin, indicating that the model is unstable and there is no way to constrain the secondary component velocity.
This behavior is explained by the fact that the secondary component is lost in the dominant stellar population and has less prominent absorption features due to its younger age.

\subsection{Ionized gas}
\label{sec:ionized_gas}

\begin{figure}
    \centering
        \includegraphics[width=0.8\columnwidth]{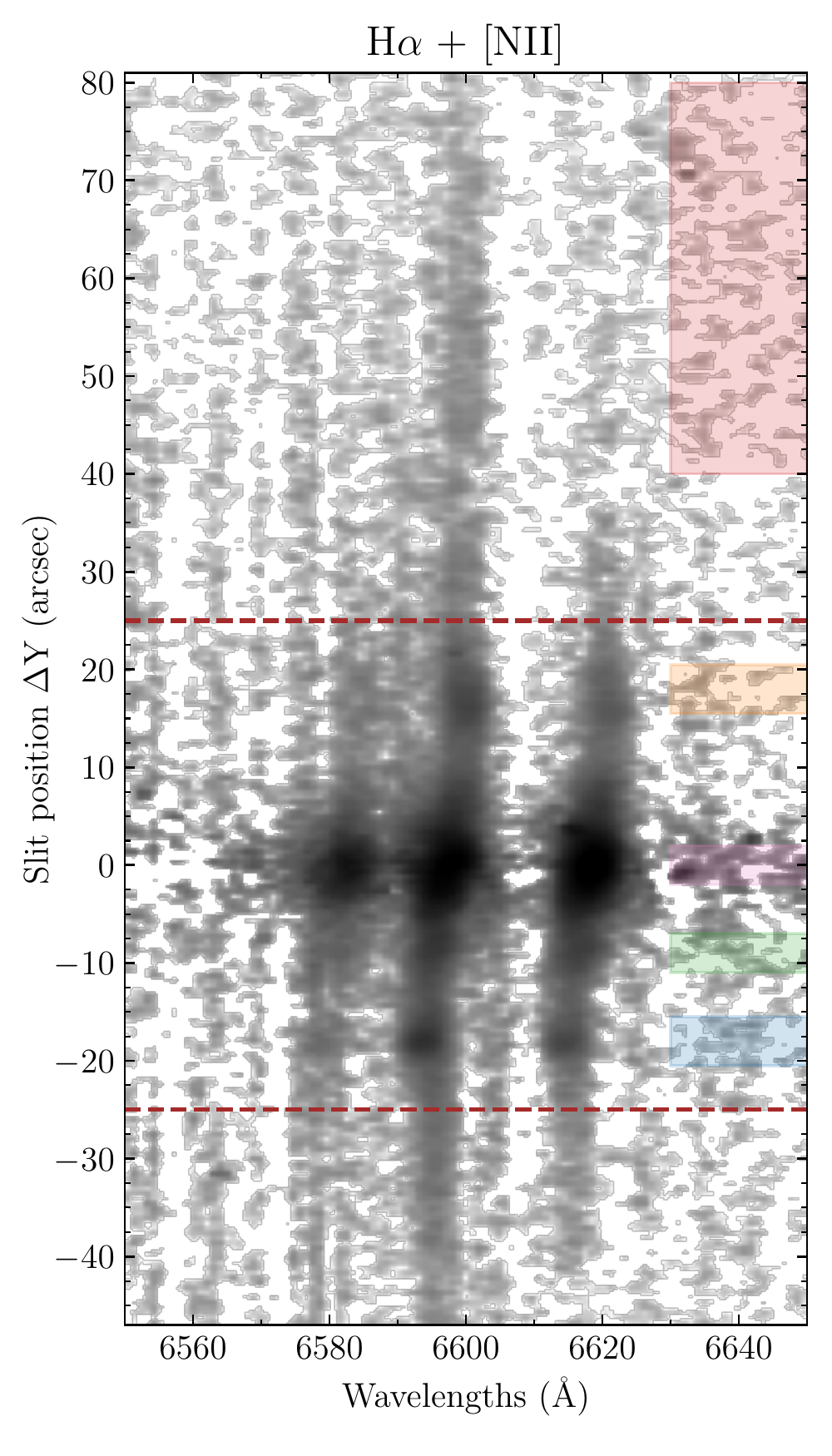}
    \caption{Emission line flux distribution along the slit. Starlight contribution is removed. Brown dashed lines show locations of the stellar inner ring $R=\pm 25$\arcsec. Color stripes correspond to the same radial ranges as shown in Fig.~\ref{fig:spresult2}.}
    \label{fig:emission}
\end{figure}

Emission lines are found across the entire disk of the galaxy.
This can be seen in Fig.~\ref{fig:emission}, where we demonstrate a fragment of the long-slit spectra around the \Ha+[N\ii] lines cleaned from the stellar continuum.
Remarkable features of the emission-line distribution are two prominent, symmetrical, clump-like brightenings at $R=\pm18\arcsec$ (see also panel \textit{e} of Fig.~\ref{fig:spresult2}), which we interpreted as an inner emission-line ring structure, much closer to the center than the stellar inner ring at $R=25\arcsec$ according to \citet{buta_cat}. 
The outer ring also includes an emission-line component, though very diffuse, but clearly seen to the north-west of the center at the radii of $45\arcsec-80\arcsec$.
The southern part of the outer ring is contaminated by two bright stars just at the major axis, so we could not measure the emission-line fluxes there.

\begin{figure*}[hbt!]
\centering
\includegraphics[width=0.85\textwidth]{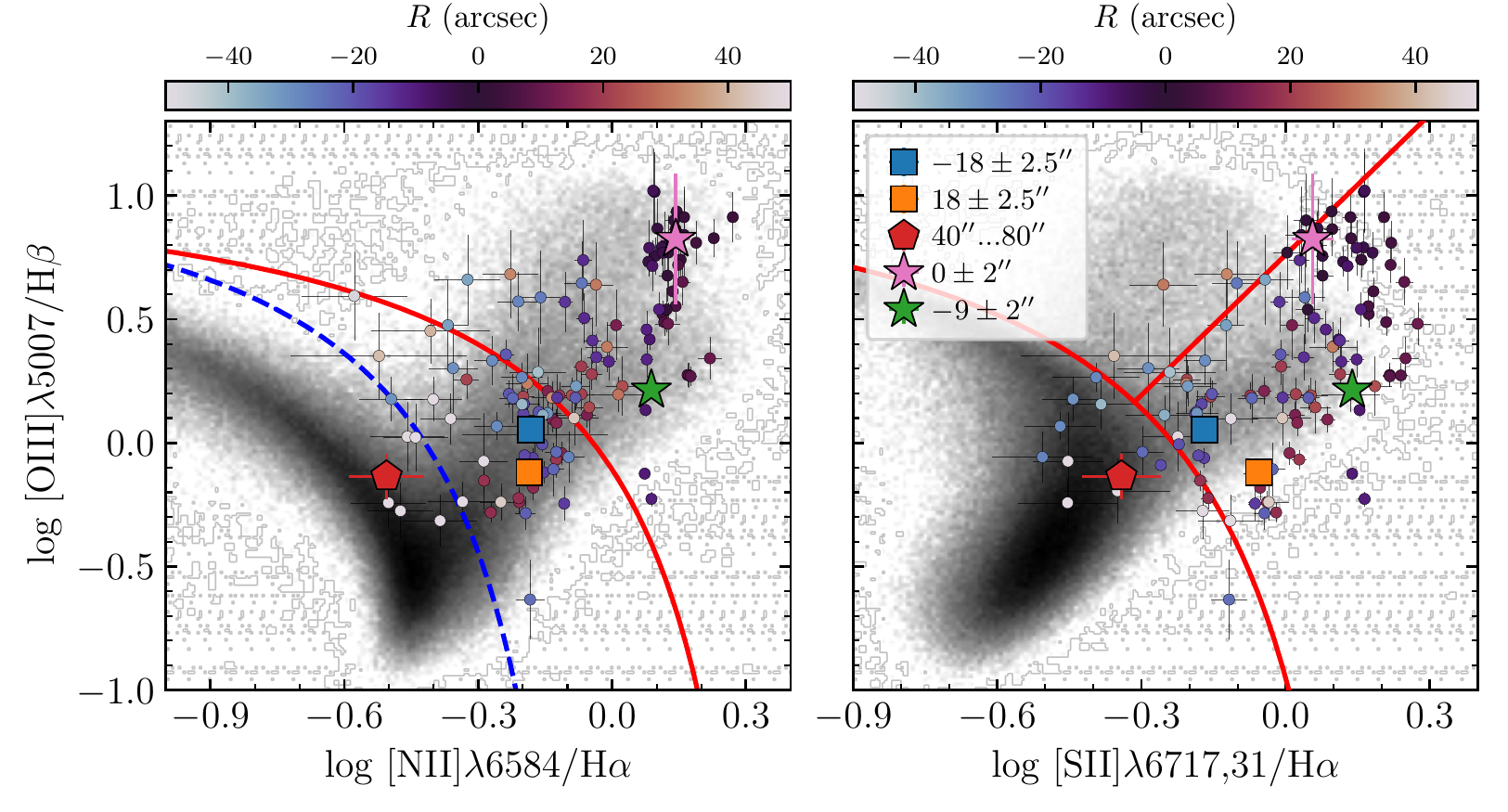}
\caption{Strong-line ratio diagnostic diagram.
Black points with error bars represent  our measurements in the individual spectra, while the colored squares refer to the average values for the inner and outer emission-line rings.
The nucleus is shown by a large purple star.
The averaged values marked by color symbols correspond to the stripes of the same color in Figs.~\ref{fig:spresult2},\ref{fig:emission}.
The red lines show theoretical demarcation lines by \citet{kewley01} separating the parameter space according to different excitation mechanisms.
The blue dashed line is an empirical separating line from \citet{kauffmann03}.
The distribution of emission-line measurements taken from the Reference Catalog of Galaxy SEDs \citep[RCSED, \url{http://rcsed.sai.msu.ru}, ][] {Chilingarian2017ApJS..228...14C} is shown in gray.}
\label{fig:bpt2}
\end{figure*}

We revealed gas excitation mechanisms by using standard Baldwin-Phillips-Terlevich diagram \citep[BPT,][]{bpt} and demarcation lines by \citet{kewley01} and \citet{kauffmann03} (see Fig.~\ref{fig:bpt2}). 
First of all, we note that the dominant excitation smoothly changes along the radius from a low-ionization by active galactic nuclei (AGN) in the center of the galaxy (dark purple symbols) to ionization by young stars on the outskirts of the galaxy (light symbols).
We expected to see HII-regions in the inner ring, which is bright in UV; however, the measurements there are located in the transition zone implying the contribution of shocks into the gas excitation.
The diffuse gas of the outer ring demonstrates pure excitation by young stars.

Thus, we can determine its oxygen abundance by using strong line calibration.
We used N2 and O3N2 calibrations proposed by \citet{pp04} and \citet{marino13}.
Both indicators from both papers have given consistent estimates for the outer ring: $12+\log \mbox{O/H} = 8.48 \pm 0.08$~dex.
For the inner ring, we only used the O3N2 calibration because it is reliable in the transition zone of the BPT-diagram \citep{kumari19}.
The oxygen abundance estimate for the inner ring is $12+\log \mbox{O/H} = 8.56 \pm 0.06$~dex.
Hence, the ionized-gas metallicity in NGC~254 is slightly subsolar (-0.2 to -0.13~dex assuming a solar abundance of 8.69 from \citet{Asplund2009ARA&A..47..481A}), with the small negative radial metallicity gradient within our errors.

While inspecting a purely emission-line spectrum, we also found noticeable emission in the neutral resonant Na I D ($\lambda5895.92$\AA, $\lambda5895.92$\AA) lines at radii $R=2\arcsec-10\arcsec$ on either side of the center of the galaxy.
Kinematically, it does not differ from other lines and demonstrates rotation in the opposite direction relatively to the stars.
Na I D lines in emission are rarely observed in galaxies and are most usually associated with the multiphase nature of the outflowing gas \citep{Rupke2015ApJ...801..126R, Perna2019A&A...623A.171P, Rupke2021MNRAS.503.4748R} and specific photoionization conditions \citep{Baron2020MNRAS.494.5396B}.
The LINER-type excitation in the center supports the idea of AGN activity, which can drive gas outflows.
Meanwhile, we did not find any noticeable kinematic signs of ionized outflows, which usually manifest themselves as non-Gaussian emission-line profile shapes and/or strong blue wings of the [O\iii] line profiles, in particular.

\section{Discussion: Star formation in NGC~254 and ring nature}
\label{sec:discussion}

The remarkable morphological appearance makes ringed galaxies attractive for investigation.
However, it is unclear how different the evolutionary paths are for ringed and non-ringed galaxies \citep{Fernandez2021A&A...653A..71F,eagle_rings}.
A comprehensive dataset and our dedicated analysis help us to build a consistent picture for the double-ringed galaxy NGC~254.
One can note from Fig.~\ref{fig:image} that the galaxy has generally disturbed outer parts, looking like asymmetric light distribution, similar to shells, that may be due to gravitational interaction with neighboring satellites or past merger event.
The same interactions can be a factor for the formation of ring structures in NGC~254 \citep{tutukov_fedorova,eliche-moral11,mapelli15}.
Another factor possibly responsible for the rings may be a non-axisymmetric potential \citep{buta_combes}.
In the central part of NGC~254, one can see such a structure, which is considered as a bar by some authors \citep{S4G_morph}.
However, other studies have argued that NGC~254 is a non-barred galaxy.
We tend to consider this central non-axisymmetric structure as an oval lens which may be a relic of the dynamic evolution of a bar \citep{Kormendy1979ApJ...227..714K}.
In any case, the presence of this non-axisymmetric structure implies the resonance nature of the inner ring.

Our emission-line analysis (see Section~\ref{sec:ionized_gas}) suggests that the ionized gas is mainly excited by young stars inside two rings of NGC~254, both the inner and the outer ones.
Despite the contamination of the gas excitation by shock contribution in the inner ring, the UV flux map and the $FUV-NUV$ distribution provide extra evidence for the presence of young stars there.
However, we need to point out that the inner star forming ring does not coincide with the inner {\it \emph{stellar}} ring -- it is closer to the center by 8\arcsec\ (0.7~kpc).
The same shift is demonstrated by the ultraviolet map constructed based on the GALEX data (Fig.~\ref{fig:spresult2}, panel \textit{b}); the $FUV-NUV$ radial profile found by \citet{S4G_galex} also reveals a minimum, $FUV-NUV=0.385\pm 0.128$, which is typical for young stars at $R=18\arcsec,$ while at $R=24\arcsec$ the UV color is already larger than 1, which is typical for older stellar populations.
Meanwhile, in the color map $g-r$ the inner stellar ring at $R=25\arcsec$ is markedly blue, while the inner emission-line (star forming) ring contains a red dust lane (Fig.~\ref{fig:spresult2}, panel \textit{c}). Such configurations, including shocks revealed by dust concentration and the BPT-diagram inside the blue inner stellar ring, imply a rapid drift of the star forming zone toward the galactic center.

By combining all our and literature data, we can make the following comments concerning global star formation in NGC~254.
It is currently proceeding, both in the outer and in the inner rings.
It is rather weak.
Currently, stars are emerging from the {\it \emph{counter-rotating}} gas.
In the inner ring, we detected a young stellar component contributing $\approx1\%$ to the stellar density (see Section~\ref{sec:spec_phot_fitting}).
We now look at another way to estimate the contribution of counter-rotating stars to NGC 254.
Only 1~Gyr ago, star formation in the disk of NGC~254 proceeded in the {\it \emph{corotating}} gas because now we see stellar population with the luminosity-weighted age of 2--3~Gyr there (Fig.~\ref{fig:spresult}) -- here we used the relation between SSP-equivalent age and quenching time from \citet{Smith2009MNRAS.392.1265S}. Hence, we can fix the epoch of counter-rotating gas accretion as 0.5--1.0~Gyr ago. The current star formation rate is 0.02~\Ms\ per year if we apply the star formation rate calibration from \citet{sfr_fuv} to the total FUV flux retrieved from NED.
Interestingly, NGC~4513, which was studied by us previously \citep{n4513_aa} and also has a counter-rotating gaseous ring, proceeds star formation at exactly the same rate.
Under the limit of flat star formation history \citep[while it must be decreasing,][]{wesaltrings}, we obtain $<2\times 10^7$~\Ms\ counter-rotating stars.
The total mass of dynamically cold stellar components, disk$+$bar, is $8\times 10^9$\Ms, according to \citet{S4G_disks}.
The counter-rotating stellar component may contribute less than 1\%\ to the total disk mass.
It agrees very well with our estimate based on the two-component spectro-photometric analysis at $|R|=18\arcsec$.

The analysis for regions beyond the radius of $|R|=18\arcsec$ shows a higher contribution from the secondary component, but it is not young enough for us to confidently associate it with stars freshly formed from counter-rotating gas.
Our non-parametrical LOSVD analysis (Section~\ref{sec:loking_for_CR}) gives some clues about the presence of counter-rotating stars at $|R|=5-15\arcsec$.
Given that we also see an increase of velocity dispersion in the outer part of the galaxy $R>+40\arcsec$ and disturbed morphology in the galaxy outskirts, we tend to consider that NGC~254 has experienced the merger event in the past.
A retrograde merger with a gas-rich galaxy in the past \citep{Thakar1996ApJ...461...55T, Thakar1998ApJ...506...93T} could naturally explain the large-scale counter-rotating gaseous disk, the disturbed morphology, the signs of counter-rotating stellar components in the central region $R<15\arcsec$ as a relic of an accreted stellar body, and the dynamically heated outer part of the disk.
We cannot completely rule out retrograde accretion from cosmological filaments as the main source of counter-rotating gas \citep{Algorry2014MNRAS.437.3596A, khoperskov_illustris}; however, in the case of NGC~254 it seems less likely, because it does not explain the morphological disturbed appearance of NGC~254.
\citet{khoperskov_illustris} studied galaxies with stellar counter-rotation in the IllustrisTNG simulations and showed how existing gas in their disks can initially be replaced by external accretion from retrograde orbits, accompanied by an efficient inflow to the center causing rejuvenation of the stellar population and AGN activity.
Although in this simulation the formation of counter-rotation is due to the retrograde gas accretion from cosmological filaments, the dynamics of the gas mixture and inflow may be similar to the case of gas-rich satellite merging. This may be applicable to NGC~254, explaining the central rejuvenation and AGN activity.

We measured the gas metallicity in the rings of NGC~254. This is $-0.13\pm 0.06$~dex in the inner ring and $-0.21\pm 0.08$~dex in the outer ring. It is fully consistent with our early results on the S0 outer starforming rings: in \citet{s0_fp}, we obtained -0.15~dex for all the outer starforming rings of our sample S0s independently of the ring's relative radius or galaxy luminosity.
We explain this uniformity in the frame of modern chemical evolution models \citep[e.g.,][]{ascasibar}, demonstrating that the gas metallicity reaches a plateau close to the solar value very soon after the star formation starts at some particular radius. This is exactly when the stellar--to--gas local density ratio exceeds a unity.
Since the ring star formation bursts in S0s having accreted the outer gas are short, with e-folding times of 600~Myr and less \citep{wesaltrings}, we almost always meet their saturated regime of chemical evolution.

\section{Conclusions}
\label{sec:conclusions}

We studied a lenticular galaxy with two rings, NGC~254. Our spectroscopy revealed the ionized gas counter-rotating with respect to the stars across the entire galaxy disk.
The gas emission confined, besides the nucleus, to the inner and outer rings is excited mostly by young stars, and the oxygen abundance of the gas estimated through the strong-line calibrations is slightly below the solar value.
The total star formation rate estimated from the $FUV$ flux and related mostly to the inner
ring is 0.02 solar mass per year.
We suggest a scenario of quite recent (about 1~Gyr ago) outer gas accretion from a retrograde orbit by a previously gas-rich disk galaxy.
As a consequence of this event, the primary corotating gas of the galaxy has lost its momentum and has fallen into the center, feeding the stellar nucleus rejuvenation and AGN activity.
The slightly disturbed morphology of the galaxy also suggests  the merger happened in the past.
Externally accreted counter-rotating gas formed two star forming rings contributing, to date, $\approx1\%$ to the local stellar density.
Their radial locations are restricted perhaps by resonances of the oval central lens of NGC~254, which may be a result of galaxy disturbance by minor merging \citep{eliche-moral11,mapelli15}.
However, the fast radial drift inward of the inner star forming ring found in our spectral data is difficult to explain in the frame of this widely accepted model.

In addition, we would like to point out that our spectro-photometric approach used for the estimation of counter-rotating stellar contribution broadens our horizons with regard to the revision of the frequency of large-scale, stellar, counter-rotation phenomena, particularly in spirals with current star formation.
An early study by \citet{Pizzella2004A&A...424..447P} showed that less than 12\%\ of their sample of spiral galaxies demonstrate gaseous counter-rotation, while stellar counter-rotation was detected in less than 8\% of the sample.
Application of the proposed spectro-photometric approach can potentially equalize these frequencies due to its sensitivity to the small contribution of newly formed stars, in contrast to the standard methods based on the detection of stellar LOSVD, which requires a prominent older secondary component \citep{Rubino2021A&A...654A..30R}.

\begin{acknowledgements}
We are very grateful to the anonymous referee for very constructive and helpful comments and suggestions that made our paper much better. 
We also thank Dr. Igor Chilingarian for the fruitful discussion of the manuscript of the paper.
The study of galactic rings was supported by the Russian Foundation for Basic
Researches, grant no. 18-02-00094a and by the National Research Foundation (NRF) of South Africa.
We also acknowledge the financial support from the Interdisciplinary Scientific and Educational School of the Lomonosov Moscow State University ``Fundamental and Applied Space Research'' (COSMOS).
IK and DG acknowledge the support from the Russian Scientific Foundation grant 19-12-00281 and 21-72-00036.
The work is based on spectral data obtained at the Southern African Large Telescope under program 2018-1-MLT-003 and on the public data of the DESI (http://legacysurvey.org) and GALEX (http://galex.stsci.edu/GR6/) surveys.
This research has made use of the services of the ESO Science Archive Facility.
Based on observations collected at the European Southern Observatory under ESO programme ID 179.A-2004.
The NASA GALEX mission data were taken from the Mikulski Archive for Space Telescopes (MAST).
The WISE data exploited by us were retrieved from the NASA/IPAC Infrared Science Archive, which is operated by the Jet Propulsion Laboratory, California Institute of Technology, under contract with the National Aeronautics and Space Administration.
The Legacy Surveys imaging of the DESI footprint is supported by the Director, Office of Science, Office of High Energy Physics of the U.S. Department of Energy under Contract No. DE-AC02-05CH1123, by the National Energy Research Scientific Computing Center, a DOE Office of Science User Facility under the same contract; and by the U.S. National Science Foundation, Division of Astronomical Sciences under Contract No. AST-0950945 to NOAO.
This research made use of Astropy\footnote{\url{http://www.astropy.org}}, a community-developed core Python package for Astronomy \citep{AstropyCollaboration2013A&A...558A..33A, AstropyCollaboration2018AJ....156..123A}.
\end{acknowledgements}

\bibliographystyle{aa} % style aa.bst
\bibliography{bib} % your references Yourfile.bib

\end{document}